\documentclass{iaus}
\usepackage{graphicx,natbib}

\bibpunct[, ]{(}{)}{;}{a}{}{,}

\def \mnras {MNRAS}
\def \apj {ApJ}

\def \aap {A\&A}

\def \logm {$\log({\rm M/M}_\odot)$}

\title[Building the red sequence] 
{Building the red sequence through gas-rich major mergers}

\author[Wild et al.]   
{Vivienne Wild$^1$, C. Jakob Walcher$^2$, Peter H. Johansson$^3$}

\affiliation{$^1$ Institut d'Astrophysique de Paris, CNRS, Universit\'{e}
Pierre \& Marie Curie, \\
UMR 7095, 98bis bd Arago, 75014 Paris, France. email: {\tt wild@iap.fr} \\[\affilskip]
$^2$European Space Agency, Keplerlaan 1, 2200AG Noordwijk, The Netherlands\\[\affilskip]
$^3$Universit\"{a}ts-Sternwarte M\"{u}nchen, Scheinerstr. 1, D-81679
M\"{u}nchen, Germany}

\pubyear{2009}
\volume{262}  
\pagerange{119--126}
\jname{Stellar populations, planning for the next decade}
\editors{A.C. Editor, B.D. Editor \& C.E. Editor, eds.}
\begin{document}

\maketitle

\begin{abstract}
  Understanding the details of how the red sequence is built is a key
  question in galaxy evolution. What are the relative roles of
  gas-rich vs. dry mergers, major vs. minor mergers or galaxy mergers
  vs. gas accretion? In a recent paper \citep{Wild:2009p2609}, we
  compare hydrodynamic simulations with observations to show how
  gas-rich major mergers result in galaxies with strong post-starburst
  spectral features, a population of galaxies easily identified in the
  real Universe using optical spectra. Using spectra from the VVDS
  deep survey with $<z>=0.7$, and a principal component analysis
  technique to provide indices with high enough SNR, we find that 40\%
  of the mass flux onto the red-sequence could enter through a strong
  post-starburst phase, and thus through gas-rich major mergers. The
  deeper samples provided by next generation galaxy redshift surveys
  will allow us to observe the primary physical processes responsible
  for the shut-down in starformation and build-up of the red sequence.
\end{abstract}

\firstsection 

\section{Introduction}
Recent observations have revealed that since a redshift of around
unity the total mass of stars living in red sequence galaxies has
increased by a factor of two \citep[e.g.][]{2004ApJ...608..752B}. At
the same time, the stellar mass density of the blue sequence has
remained almost constant. The interpretation is that some blue
galaxies migrate onto the red sequence after the quenching of their
star formation, whilst the remainder continue to form new stars
\citep[e.g.][]{2007ApJ...665..265F}.

\citet{2007A&A...476..137A} measured the net mass flux which has taken
place from the blue sequence to the red sequence. This amounts to
$9.8\times10^{-3}$M$_\odot$/yr/Mpc$^3$, or about
$1.4\times10^4$\,M$_\odot$/yr in the VIMOS VLT DEEP Survey (VVDS)
volume. Recent star formation history can be used as a tool to
identify galaxies as they enter the red sequence. For typical
observable times of post-starburst features in VVDS optical galaxy
spectra of $\sim0.35-0.6$\,Gyr, this could comprise, for example, a
few tens of galaxies of stellar mass \logm$=10.5$ in the VVDS survey.
In these proceedings I will summarise the work presented in detail in
\citet{Wild:2009p2609} and ask whether observations of post-starburst
galaxies are consistent with the growth of the red-sequence being
caused by gas-rich major mergers.

\section{Post-starbursts in major merger simulations}
\begin{figure}
\begin{center}
\begin{minipage}{\textwidth}
  \begin{minipage}{\textwidth}
    \hspace{0.5cm}
    \includegraphics[width=3.3cm]{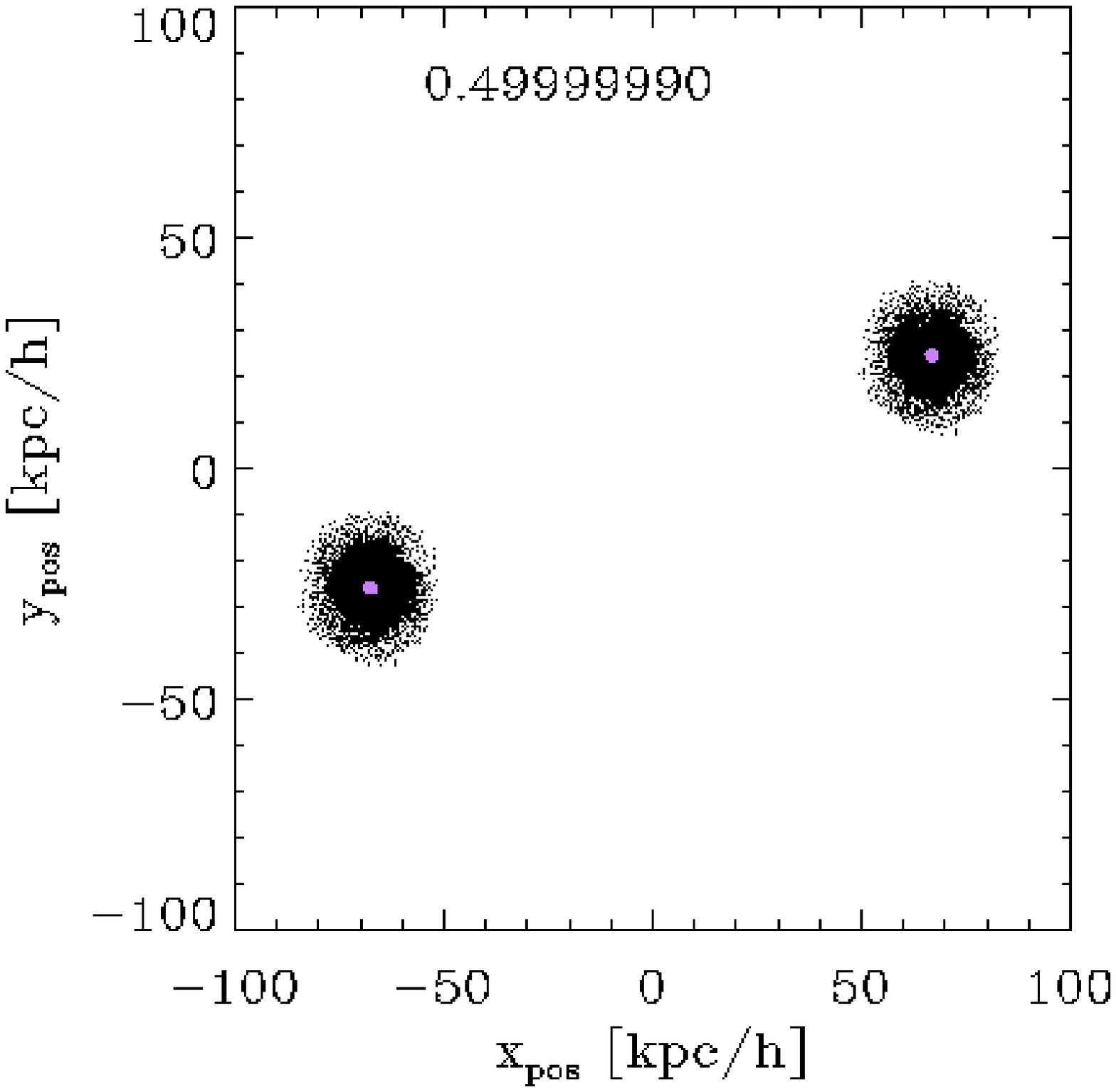}
        \hspace*{-0.4cm}
    \includegraphics[width=3.3cm]{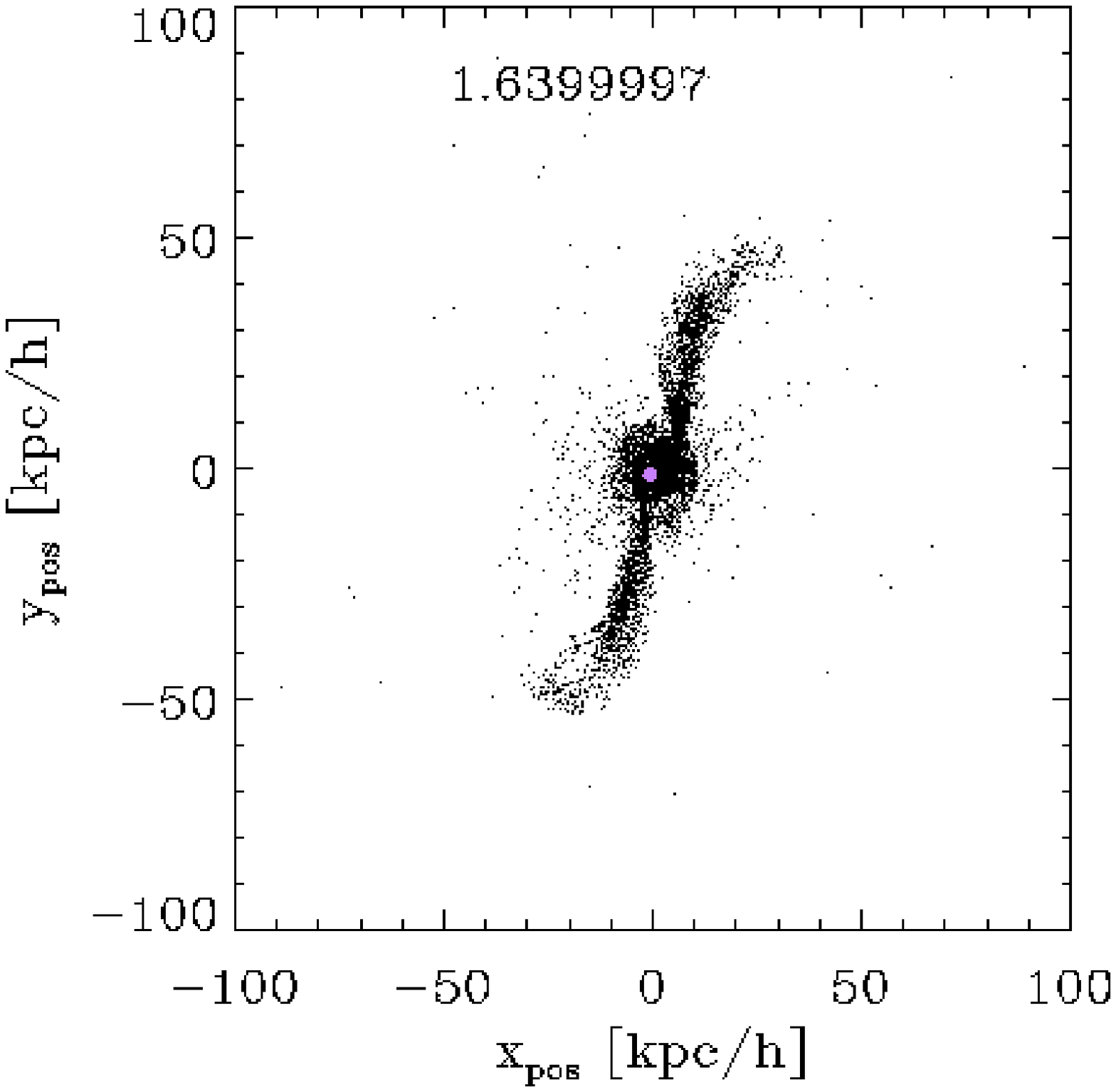}
        \hspace*{-0.4cm}
    \includegraphics[width=3.3cm]{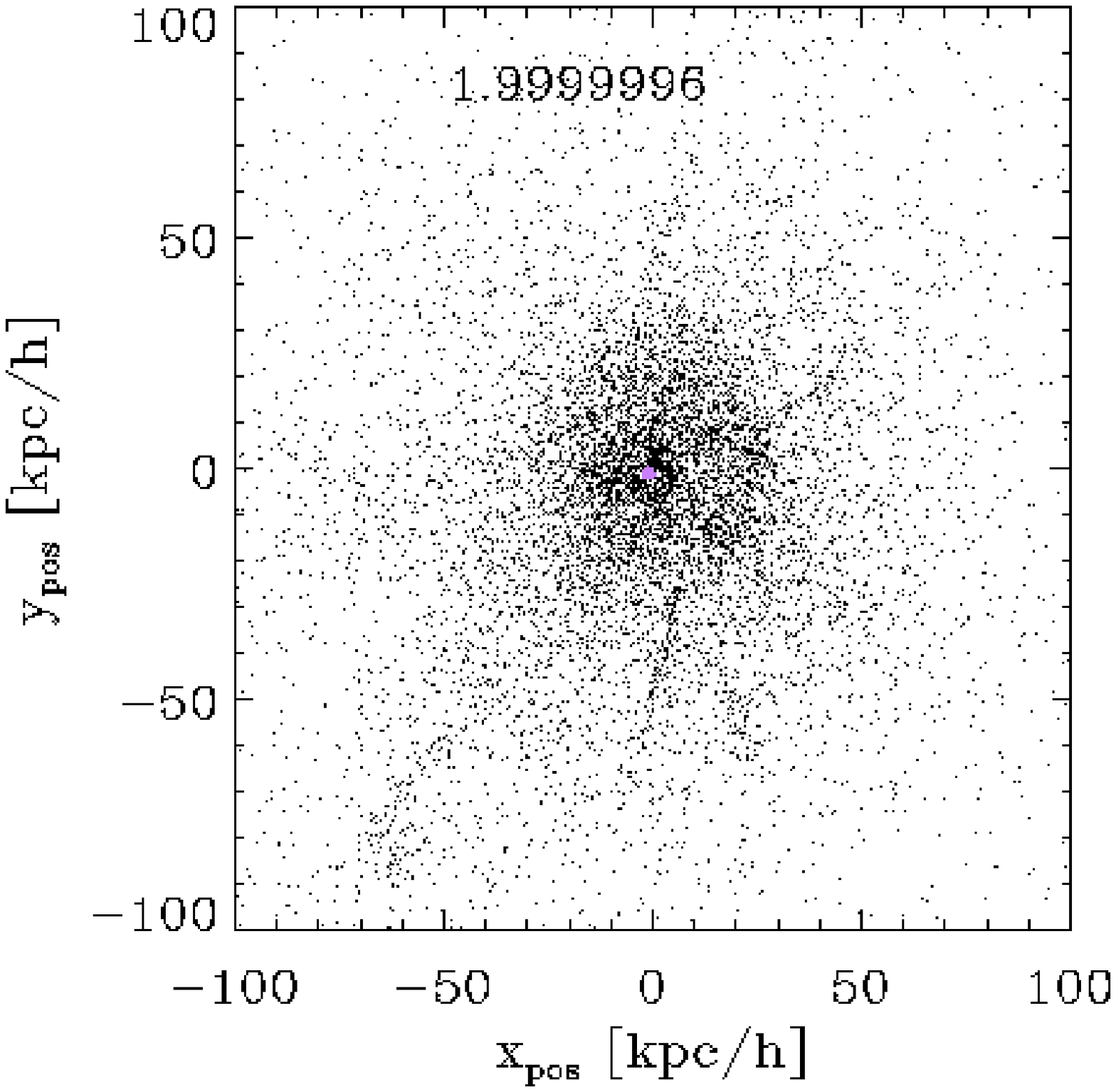}
        \hspace*{-0.4cm}
    \includegraphics[width=3.3cm]{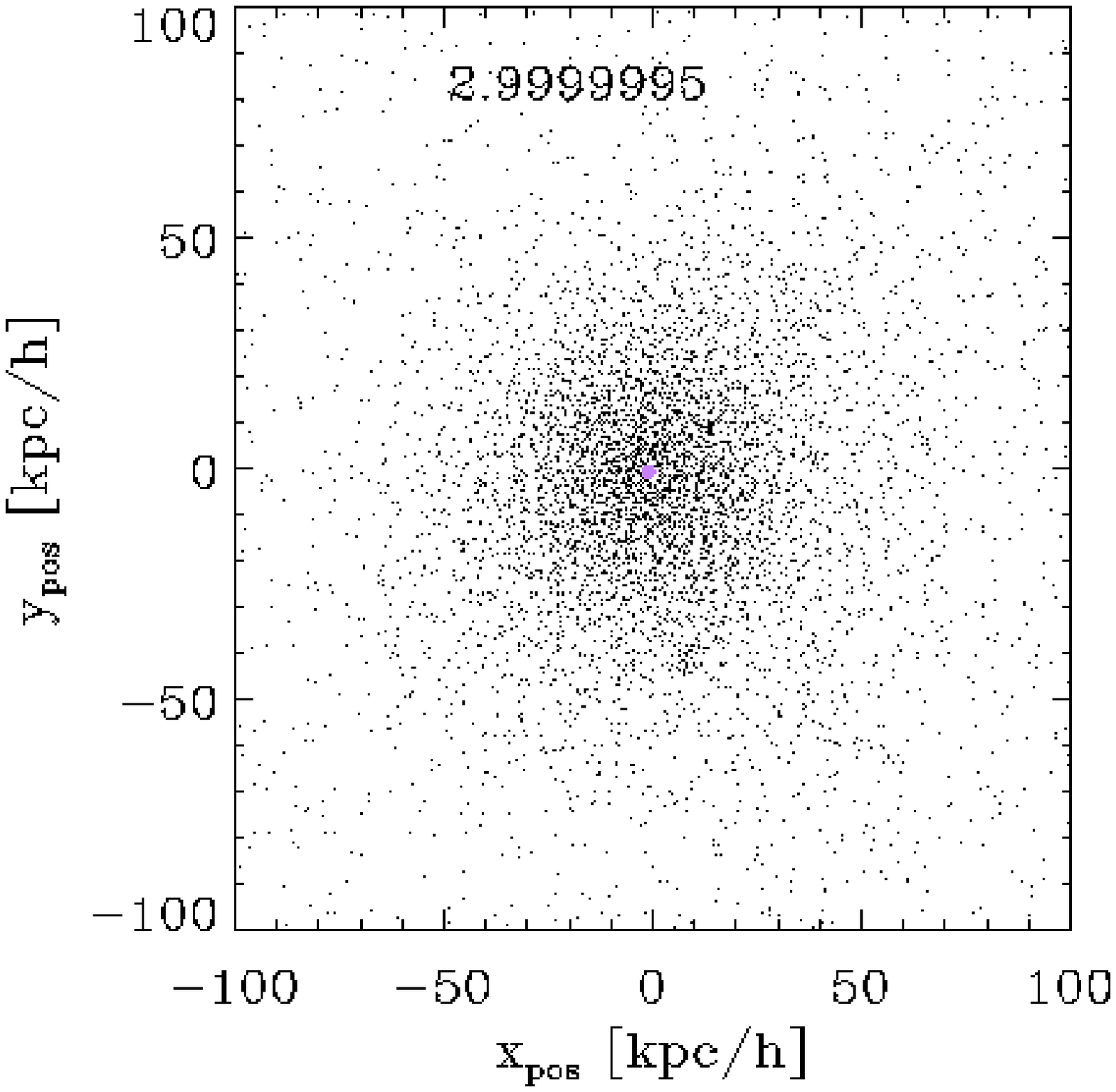}
  \end{minipage}
  \begin{minipage}{\textwidth}
    \includegraphics[width=\textwidth]{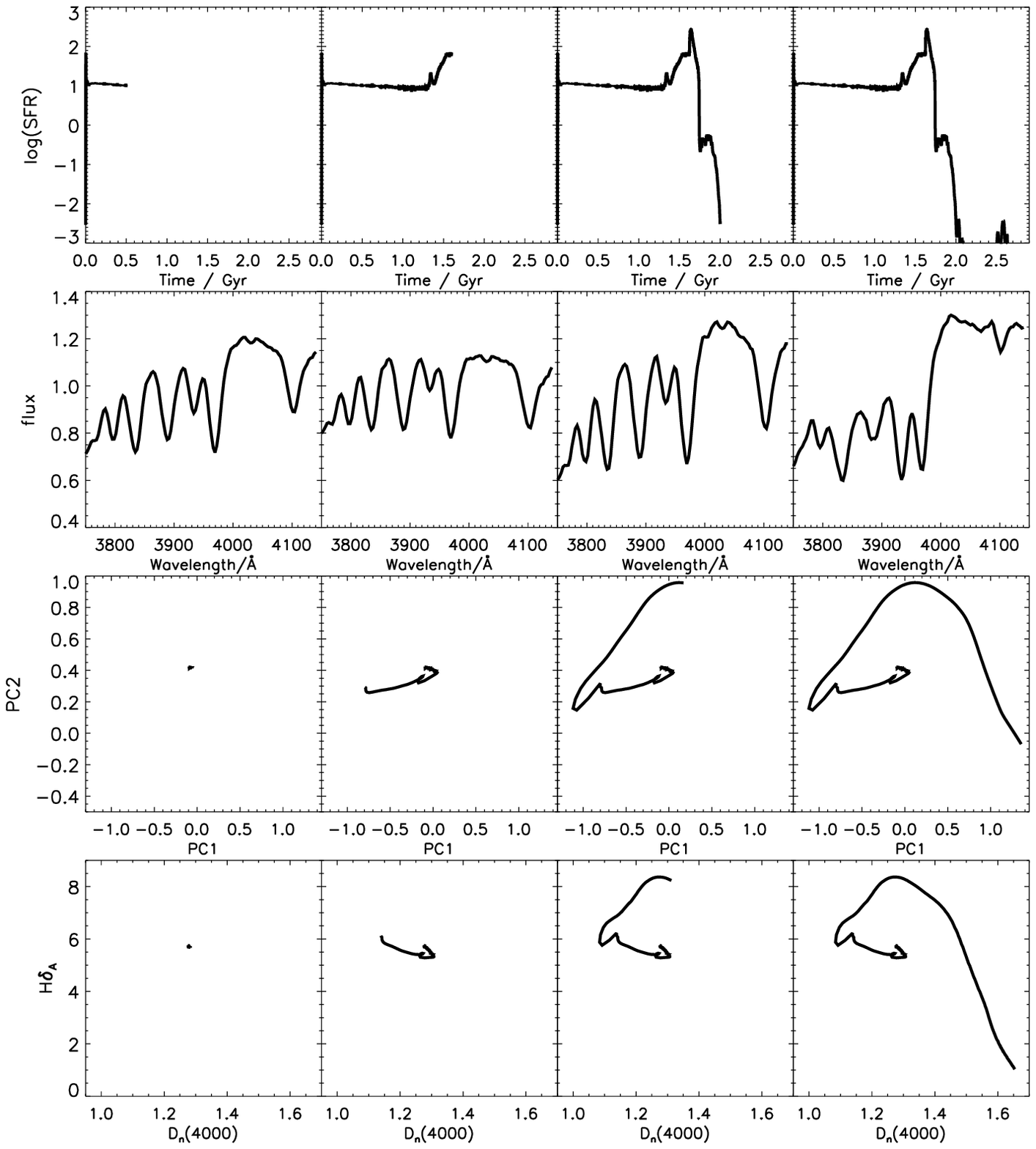}
  \end{minipage}
\end{minipage}
\caption{From left to right snapshots of a hydrodynamic simulation of
  a major galaxy merger, at 0.5\,Gyr (initial starforming disks),
  1.64\,Gyr (peak of starburst), 2\,Gyr (peak of post-starburst
  features) and 3\,Gyr (remnant) after the start of the
  simulation. From top to bottom: the $x-y$ positions of gas
  particles and black holes (purple points); trace of combined star-formation rate of both galaxies;
  the optical continuum spectrum in 4000\AA\ break region (note that
  emission lines, have not been plotted to improve clarity); the trace
  of PCA derived indices that describe the strength of the 4000\AA\
  break and the Balmer absorption lines; the trace of two indices used
  traditionally to infer the recent star formation history of
  galaxies. }\label{fig1}
\end{center}
\end{figure}

Alongside toy models of star formation histories, modern advanced
simulations afford a new means of interpretation of observational
results. Smoothed particle hydrodynamic (SPH) simulations of galaxy
mergers are now carried out routinely by several groups, and while we
await new advances in combining radiative transfer with SPH merger and
cosmological simulations \citep{2006ApJ...637..255J}, simple
comparisons can be already be made.  For a suite of 79 merger
simulations presented in \citet{2008arXiv0802.0210J} we used standard
stellar population synthesis, dust attenuation law and Balmer emission
predictions to recreate the ``observed'' spectra of the merging
galaxies. 

In Figure 1 we present four snapshots in time of one particular merger
simulation, chosen to illustrate the peak differences in the optical
spectral features. One traditional index used for identifying
post-starburst galaxies, the Balmer H$\delta_A$ absorption line is
compared with our new method which combines all 5 Balmer lines visible
in the 4000\AA\ break region, together with the changing shape of the
continuum using a Principal Component Analysis based on the method
described in \citet{wild_psb}. As the merger commences, the sudden
presence of hot O and B stars cause a rapid increase in the blue-UV
continuum causing a decrease in the measured 4000\AA\ break
strength. Within $\sim$0.5\,Gyr a combination of exhaustion of the gas
supply and disruption and expulsion of gas by the energy of many
supernovae causes the star formation rate to decline rapidly. The loss
of the contribution of short-lived O and B stars to the integrated
spectrum, and the dominance of slightly cooler and longer-lived A and
F stars, cause the strong Balmer absorption features of the
post-starburst phase. In this particular simulation, QSO feedback is
included, causing the further shut-down of star formation. However,
the presence or absence of QSO feedback makes very little difference
to the strength of the post-starburst features, or the observational
lifetime of the post-starburst phase.

From our analysis of both toy-models and merger simulations, we find
that gas-rich spiral mergers can lead to galaxies with strong Balmer
absorption features before the galaxy enters the red
sequence. However, the timescale of the decay in star formation must
be short ($<10^8$yrs) and the burst mass fraction must be large
($>5-10$\%) for post-starbursts to be observed using VVDS spectra. The
strongest starbursts visible in the VVDS dataset are detectable for
0.6\,Gyr in the post-starburst phase.

\section{VVDS post-starburst galaxies}

The VVDS is a deep spectroscopic redshift survey, targeting objects
with apparent magnitudes in the range of $17.5 < I_{AB} < 24$.  The
survey is unique for high-redshift galaxy surveys in having applied no
further colour cuts to minimize contamination from stars, yielding a
particularly simple selection function.  The spectra have a resolution
(R) of 227 and a useful observed frame wavelength range, for our
purposes, of 5500-8500\AA. In this work we select 1246 galaxies with
$0.5<z<1.0$, $I_{AB}<23$ and per-pixel S/N $>6$.

In the left panel of Figure 2 the distribution of spectral indices
``PC1'' (equivalent to 4000\AA\ break strength) and ``PC2'' (excess
Balmer absorption) is shown. The majority of galaxies form a well
defined sequence from red and dead galaxies with strong 4000\AA\
breaks to blue, star-forming galaxies with weaker breaks. A tail of
starburst galaxies exists to the bottom left, where there has been a
sharp increase in the galaxy's star formation rate over a timescale
that is short ($\sim10^8$yrs) in comparison to the age of the galaxy
(thus distinguishing them from more ordinary ``starforming''
galaxies). The orange crosses indicate the galaxies with strong Balmer
absorption lines, indicative of a recent sharp shut-down in
starformation. The encircled points indicate those galaxies with no
ongoing star formation, as measured from a fit to their
multi-wavelength spectral energy distribution
\citep{2008arXiv0807.4636W}. For comparison to other E+A samples,
these galaxies also have no visible [OII]$\lambda$3727,3730 emission,
we note that they have a tendency to be older than the star-forming
post-starbursts. In the right hand panel of Figure 2, the mass
distribution of the different types of galaxies is plotted, both
before and after correction for survey incompleteness effects. The
post-starburst sample is complete above a mass limit of
$5.6\times10^9{\rm M_\odot}$ and the non-starforming post-starbursts
have masses in the same range as red-sequence galaxies at this
redshift.

\begin{figure}
\includegraphics[width=7cm]{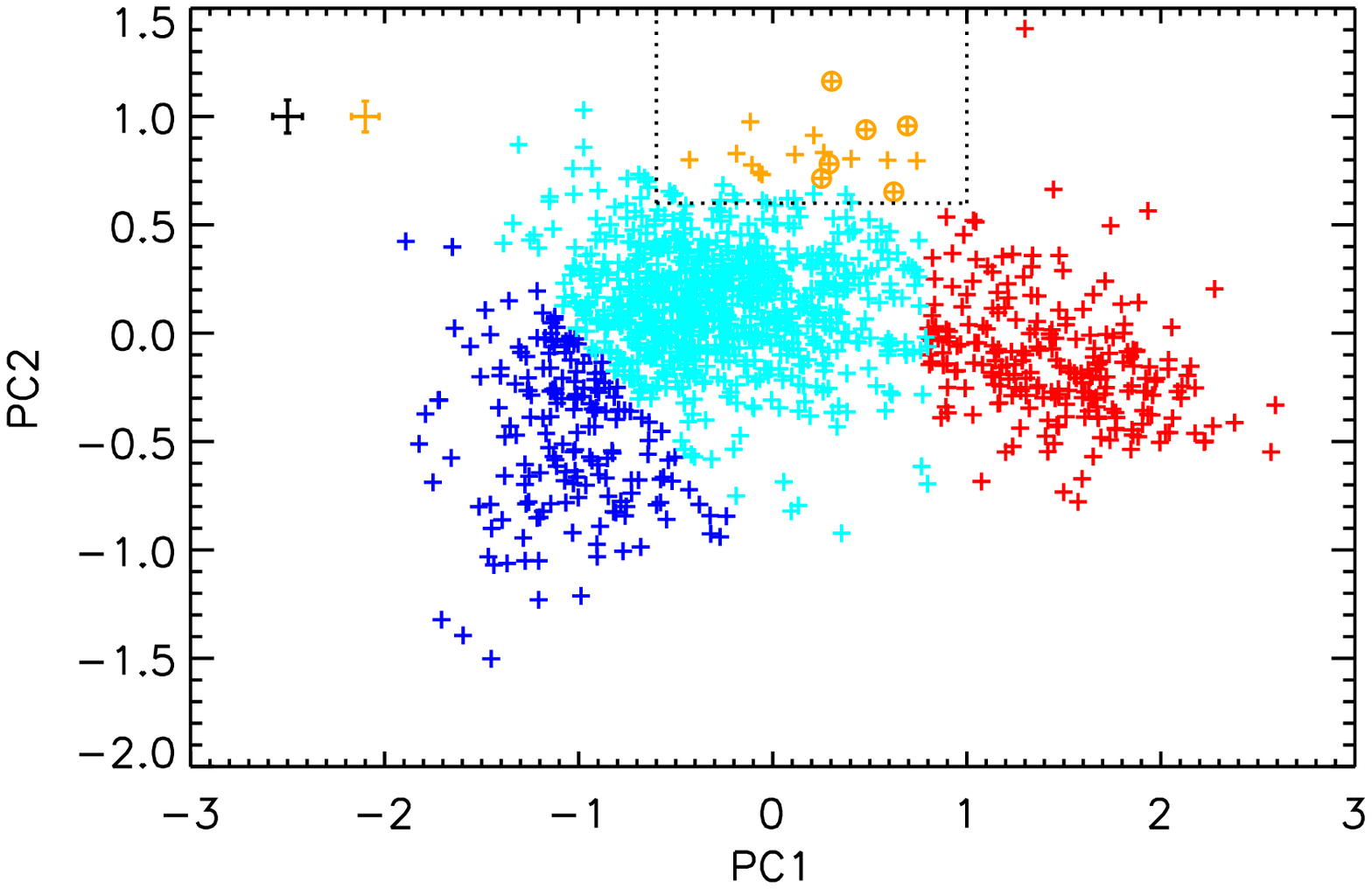}
\includegraphics[width=7cm]{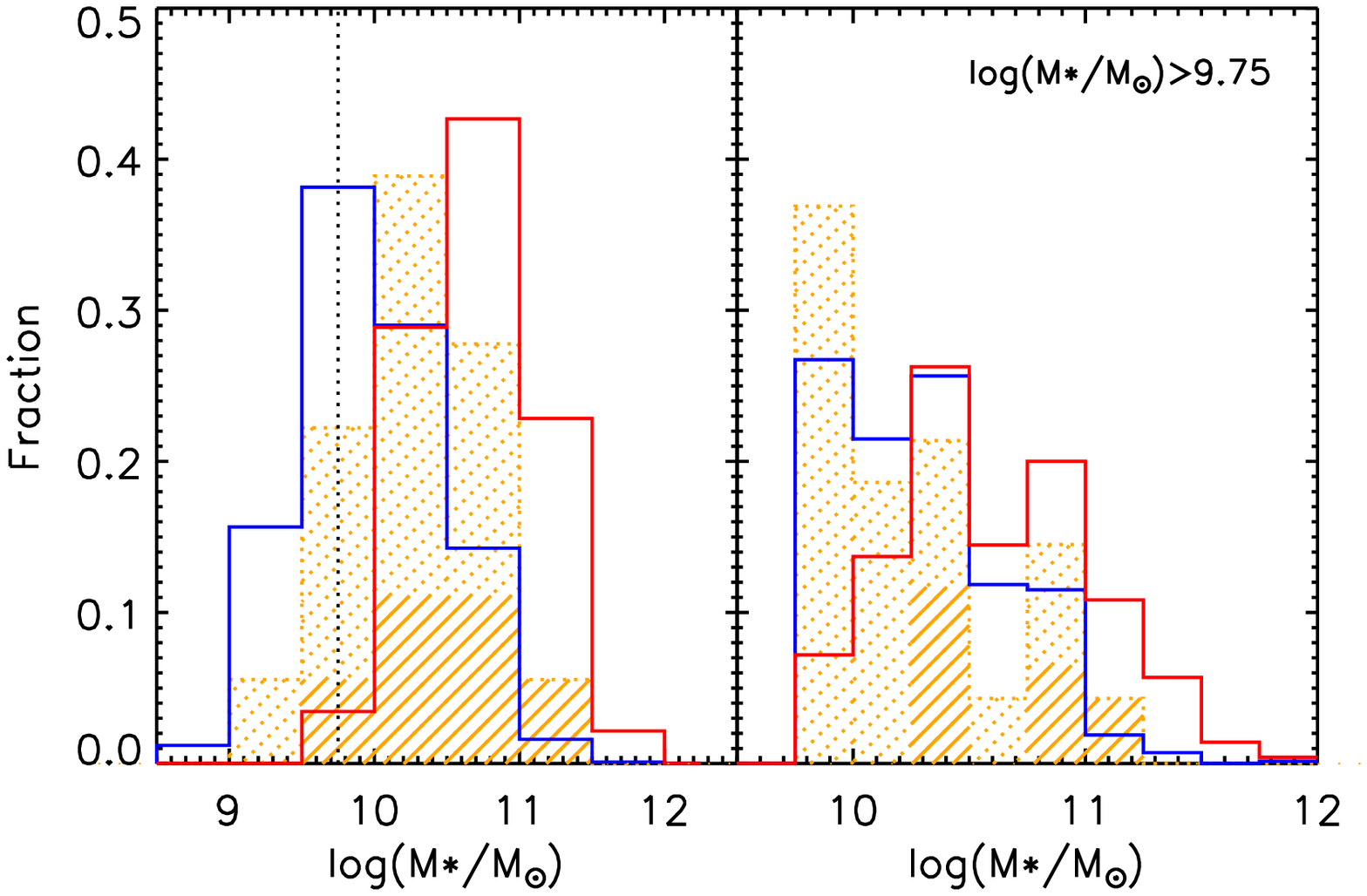}
\caption{{\it Left:} the distribution of PC1 vs. PC2 for all galaxies
  in our VVDS sample. PC1 is equivalent to the well-known index
  D$_n$4000, PC2 represents the excess (or lack) of Balmer
  absorption. For illustration purposes, the sample has been split
  into quiescent (right, red), star forming (center, cyan),
  star-bursting (bottom left, blue) and PSB (in box, orange)
  classes. Those PSB galaxies with SSFR$<10^{-11}$/yr are
  circled. {\it Right:} the mass distribution of galaxies, before
  (left) and after (right) correction for survey incompleteness
  effects. Histograms are normalised to unity. The orange dot-filled
  histograms are all PSB galaxies, the line-filled histograms are the
  subset with no ongoing residual star formation. The vertical dotted
  line indicates our PSB mass completeness limit.  }\label{fig2}
\end{figure}

\section{The mass flux onto the red sequence}

By summing the total mass of post-starburst galaxies and using our
knowledge from simulations to estimate how long they will be visible
for, we can measure the mass flux from the blue to the red-sequence
that passes through a post-starburst phase. In this calculation, we
include only the five galaxies which have completely stopped star-formation and
therefore we argue are certain to be heading for the red-sequence. 
\begin{equation}
\dot{\rho}_{\rm B\rightarrow R,PSB} = \frac{\rm M_{B\rightarrow R,PSB}}{\rm Vol\times
  t_{PSB}} = 0.0038 {\rm M_\odot/Mpc^3/yr}
\end{equation}
Comparing to the total mass flux from the blue to the red sequence
measured by \citet{2007A&A...476..137A}, this represents
$38^{+4}_{-11}$\% of the build-up of mass onto the red sequence. We
therefore conclude that post-starburst galaxies are more than an
interesting curiosity, and while this result may be regarded as
preliminary as it relies on only 5 galaxies, our study shows how
deeper spectroscopic surveys will reveal directly the physical
processes responsible for the shut-down in star-formation.


\begin{thebibliography}

\bibitem[\protect\citeauthoryear{Arnouts, Walcher, F{\`e}vre \& et al}{Arnouts
  et~al.}{2007}]{2007A&A...476..137A}
Arnouts S.,  Walcher C.~J.,  F{\`e}vre O.~L., et al. 2007, \aap, 476, 137

\bibitem[\protect\citeauthoryear{Bell, Wolf, Meisenheimer, Rix, Borch, Dye,
  Kleinheinrich, Wisotzki \& McIntosh}{Bell et~al.}{2004}]{2004ApJ...608..752B}
Bell E.~F.,  Wolf C.,  Meisenheimer K.,  et al.  2004, \apj, 608, 752

\bibitem[\protect\citeauthoryear{Faber, Willmer, Wolf \& et al}{Faber
  et~al.}{2007}]{2007ApJ...665..265F}
Faber S.~M.,  Willmer C.~N.~A.,  Wolf C., et al. 2007, \apj, 665, 265

\bibitem[\protect\citeauthoryear{Johansson, Naab \& Burkert}{Johansson
  et~al.}{2009}]{2008arXiv0802.0210J}
Johansson P.~H.,  Naab T.,    Burkert A.,  2009, \apj, 690, 802

\bibitem[\protect\citeauthoryear{Jonsson, Cox, Primack \& Somerville}{Jonsson
  et~al.}{2006}]{2006ApJ...637..255J}
Jonsson P.,  Cox T.~J.,  Primack J.~R.,  Somerville R.~S.,  2006, \apj, 637,
  255

\bibitem[\protect\citeauthoryear{Walcher, Lamareille, Vergani \& et al}{Walcher
  et~al.}{2008}]{2008arXiv0807.4636W}
Walcher C.~J.,  Lamareille F.,  Vergani D., et~al. 2008, \aap, 491, 713

\bibitem[\protect\citeauthoryear{Wild, Kauffmann, Heckman, Charlot, Lemson,
  Brinchmann, Reichard \& Pasquali}{Wild et~al.}{2007}]{wild_psb}
Wild V.,  Kauffmann G.,  Heckman T.,  et~al.  2007, \mnras, 381, 543

\bibitem[\protect\citeauthoryear{Wild, Walcher, Johansson, Tresse, Charlot,
  Pollo, F{\`e}vre \& de Ravel}{Wild et~al.}{2009}]{Wild:2009p2609}
Wild V.,  Walcher C.~J.,  Johansson P.~H.,  et~al.,  2009, \mnras, 395, 144
\end{thebibliography}

\end{document}